\documentstyle[aps,epsf,multicol,bbm]{revtex}

\newcommand{\bea}{\begin{eqnarray}} 
\newcommand{\eea}{\end{eqnarray}} 
\newcommand{\be}{\begin{equation}} 
\newcommand{\ee}{\end{equation}} 

\begin{document}
\title{Antipersistent binary time series}
\author{Richard Metzler}
\address{Institut f\"ur Theoretische Physik und Astrophysik,
Universit\"at W\"urzburg,
Am Hubland, D-97074 W\"urzburg, Germany} 
\maketitle

\begin{abstract}
Completely antipersistent binary time series are sequences in which
every time that an $N$-bit string $\mu$ appears, the
sequence is continued with a different bit than at the last
occurrence of $\mu$. This dynamics is phrased in terms of 
a walk on a DeBruijn graph, and properties of transients and
cycles are studied. The predictability of the generated time
series for an observer who sees a longer or shorter time
window is investigated also for sequences that are not
completely antipersistent.
\end{abstract}

\begin{multicols}{2}
\section{Introduction}
The analysis and generation of time series has been of interest to 
physicists in different fields: it yields insight into 
the dynamics of chaotic systems
\cite{Takens:Detecting,Sauer:Embedology} 
and is used to study such diverse systems as 
the climate and the heartbeat \cite{Bunde:Langzeit}, 
linguistics \cite{Kanter:Markov},
and, in recent times, financial markets 
\cite{Johnson:Application,Econophys}.
The prediction of time series generated by complex 
physical systems can be of immense importance, 
as evidenced by the efforts put into improving the 
weather forecast. One of the most common aspects
of time series is long-term memory, or persistence.
This paper, however, will deal with the opposite
behaviour, namely antipersistence. 

Antipersistence is the tendency of a time series to show,
at one point, the opposite behaviour than at some point in
the past. This concept is usually applied to continuous
time series \cite{Kantz:Nonlinear}, but has received 
little attention due to the dearth of natural phenomena
that exhibit antipersistence 
(for an exception, see \cite{Ausloos:Power}).
 
In this paper, antipersistence in binary time series is defined
as follows: a sequence of bits is antipersistent on
a scale of $N$ if a string $\mu$ of $N$ bits length is likely to be
followed by the opposite bit than the last time that $\mu$
appeared. This property was observed, for example, in some
parameter regimes of the conventional Minority Game 
\cite{Marsili:Trading,Savit:Comp.} and in various
alternative models based on the idea of the Minority Game 
\cite{Reents:Stochastic}. 

This paper will first deal with completely antipersistent
time series from the point of view of discrete dynamical
systems. Properties of cycles and transients will be
studied in Sections \ref{AP-propcyc} and \ref{AP-proptrans},
respectively. Concepts and results from graph theory turn out to
yield much insight.

In Section \ref{AP-time}, the 
sequence is treated as an anti-predictable sequence for a special 
prediction algorithm, and antipersistence on one time
scale is related to properties on other time scales. Then a
more general stochastic model is introduced, and the
effects of stochasticity on the previous results are discussed.

\section{The Model}
\label{AP-model}
\begin{figure}
  \epsfxsize= 1.0\columnwidth
  \epsffile{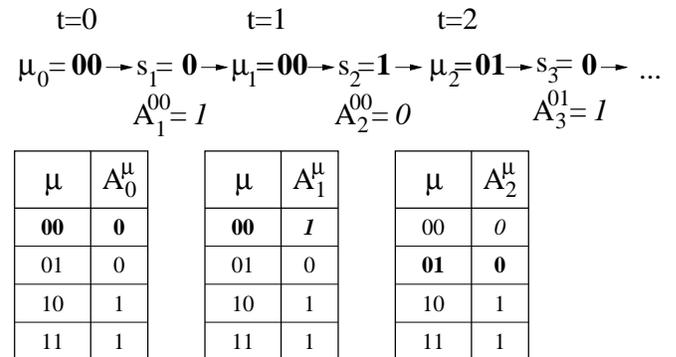}
  \caption{An example of a decision table with $N=2$ and two
steps of the dynamics. Boldface numbers indicate the current
history and the table entry used for continuing the
sequence;
italic numbers denote the last table entry that was changed.
The sequence generated in this example is $010\dots$.} 
  \label{AP-Tab}
\end{figure}

An infinitely long, completely antipersistent binary time 
series of bits $s_{\tau}\in \{0,1\}$ can by definition be generated in
the following fashion: 
at time $t$, the history $\mu_t$ denotes the binary
representation of the last $N$ bits
$s_{t-N+1},\dots,s_{t}$. 
For each of the $2^N$ possible histories $\mu$ 
there is an entry $a^{\mu}_t \in \{0,1\}$ in a decision 
table $A_t$, which also depends on $t$.
At each time step, 
\begin{itemize}
\item a new bit is generated by taking the 
table entry corresponding to the current history $\mu_t$
(which I will also refer to as ``the pattern''):  
$s_{t+1} = a^{\mu_t}_t$;
\item the history is updated: $\mu_{t+1} = (2\mu_t +
  s_{t+1}) \bmod 2^N$; i.e., all bits are shifted 
  one position to the
  left (multiplication with 2), the newly generated bit is
  added, and the oldest (most significant) bit is dropped (division
  modulo $2^N$);
\item the table entry $a^{\mu_t}_t$ that was used for making the 
decision is changed, such that the sequence will be
continued with the opposite decision when the pattern
$\mu_t$ occurs the next time: 
$a^{\mu_t}_{t+1}= 1-a^{\mu_t}_t$.
All other entries remain unchanged.
\end{itemize}
This last point is especially important. It means that the 
table entries are dynamical variables, and the state of the
dynamical system is determined by the current pattern
{\em and} the state of the decision table. Fig. \ref{AP-Tab}
shows two steps of the described dynamics of such a system
for $N=2$.

\begin{figure}
  \epsfxsize= 1.0\columnwidth
  \epsffile{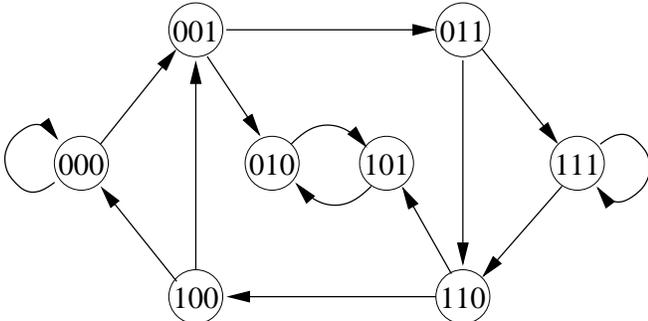}
  \caption{The directed DeBruijn graph of order 3:
Nodes represent binary strings of length $3$, edges 
lead to strings that are generated by shifting the
current string one position and adding either 0 or 1 as the
new least significant bit.} 
  \label{AP-DB}
\end{figure}

The model can also be considered from a graph-theoretical 
perspective: each pattern $\mu$ corresponds to a node on a 
directed DeBruijn graph of order $N$ (see Fig. \ref{AP-DB}
for an example of such a graph). Each node obviously 
has two edges entering it, coming from the two possible
predecessors, which I denote $^0\mu$ and $^1\mu$. 
For example, if $\mu=1100$, the possible predecessors are
$^0\mu=0110$ and $^1\mu = 1110$.
Each edge also has two
outgoing edges, leading to the two possible successors
$\mu^0$ and $\mu^1$ (for $\mu= 1100$, the successors are
$\mu^0 = 1000$ and $\mu^1 = 1001$). The graph is connected, 
since one can reach each node from any other node in a maximum of $N$ 
steps by taking the appropriate exit edges. 

In our model, at any time, only one 
of the exits leaving each node is labelled ``active'' --
the one corresponding to the table entry $a^{\mu}_t$. A time 
step consists of travelling from the current node to the
next along the active exit, then ``burning the bridge'',
i.e. labelling the previously active exit inactive and vice
versa.

\begin{figure}
  \epsfxsize= 1.0\columnwidth
  \epsffile{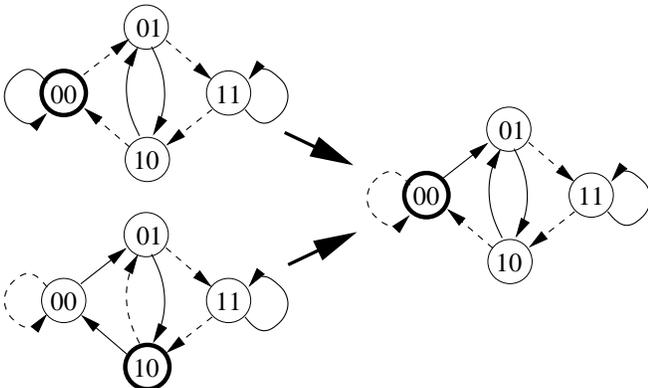}
  \caption{Example for an irreversible situation on a graph
    of order 2. Active exits are denoted by solid lines,
    inactive ones by dashed lines. The bold circle indicates
    the node currently visited. Both the upper left
    configuration (which happens to be part of a cycle, and
    corresponds to the example in Fig. \ref{AP-Tab} ) and
    the lower one (which is part of a transient) 
    lead to the configuration on the right.} 
  \label{AP-DB2}
\end{figure}

\section{Properties of cycles}
\label{AP-propcyc}
The introduced dynamics is deterministic, and the
combined system of pattern and table has a finite number  
$\Omega = 2^N \cdot 2^{(2^N)}$ of different
states, so the dynamics necessarily leads into a cycle
eventually. The dynamics is irreversible: if a 
currently visited node has two inactive entrances, it is
impossible to tell which path the system took to get to its
current state (for an example, see Fig. \ref{AP-DB2}).
 This means that not every state can be part
of a cycle, so we will have to consider the necessary
conditions for being in a cycle. I will show, step by step, 
that all cycles are of length $2\cdot 2^N$ and touch all nodes
exactly twice. 
 
Some of the proofs that now follow are redundant; on the
other hand, they help to understand the properties of the
system, and some of them are applicable to generalizations
of the problem, whereas others are not.

Let us assume that at time $0$, the system is already moving
on a cycle of length $l$. We count the number of times that
a history $\mu$
has occurred between time $0$ and time $t$ by a visit number 
$v^{\mu}_t$. Since the definition of a cycle is that after
$l$ steps the system must be in the same state again, it is 
necessary that {\em $v^{\mu}_l$  is even for all $\mu$},
since the table entries $a^{\mu}$ return to their original
state only after even numbers of visits.

Also, {\em all possible nodes are part of the cycle}. Let us prove
this by assuming the opposite, namely that there are some 
nodes that are not touched by the cycle. Since the graph is 
connected, there must be unused connections between the 
part of the graph involved in the cycle and the part that is
left out. But as we have seen in the paragraph before, the
visit number of each of the nodes that are actually part of
the cycle must be at least
2 (larger than 0, and even), so each of its two exits is 
used, including the one leading to the part of the graph 
supposedly not included in the cycle. This is a
contradiction, so all nodes are involved.

An even stronger statement is possible: the total number of visits to the
predecessors $^0\mu$ and $^1\mu$ of $\mu$ must be equal to 
twice the number of visits to $\mu$, since exactly half of the
visits they get are followed by $\mu$, while the other 
half is followed by the so-called conjugate state
$\bar{\mu}$ of $\mu$. (For example, 001 is the conjugate
state of 000.) Thus, we have
\be
v^{\mu} = (v^{^0\mu} + v^{^1\mu})/2 \mbox{~~ for all }\mu.
\label{AP-visnum1}
\ee
This can be written as a linear equation for an eigenvector
with eigenvalue 1 of a matrix $M$, with entries 
$a_{\nu\mu} = 1/2$ if $\mu$ is a possible successor of
$\nu$ and $a_{\nu\mu}=0$ otherwise. 
For example, for $N=2$, the set of Eqs. (\ref{AP-visnum1}) looks as
follows:
\bea
\left [ \left ( \begin{array}{cccc}
    \frac{1}{2}~~ & 0~~ & \frac{1}{2}~~ & 0 \\
    \frac{1}{2}~~ & 0~~ & \frac{1}{2}~~ & 0\\
    0~~ & \frac{1}{2}~~ &  0~~ & \frac{1}{2} \\
     0~~ & \frac{1}{2}~~ &  0~~ & \frac{1}{2}
    \end{array}
\right) - \mathbbm{1}_{4} \right ] 
\left (\begin{array}{c}
v^{00}_l\\
v^{01}_l\\
v^{10}_l\\
v^{11}_l
\end{array} \right ) = \left ( \begin{array}{c} 
0\\
0\\
0\\
0\end{array}\right). 
\eea
Since the sum of columns in matrix $M$ is always 1, and the
individual entries are $\geq 0$, and it describes
transitions on a connected graph, we can apply results from
the theory of stochastic matrices to state that it has one 
unique eigenvector with eigenvalue 1 \cite{Feller},
and we easily guess that $v_l^{\mu} = \mbox{ const}$
fulfills Eq. (\ref{AP-visnum1}). That means that in a cycle,
{\em all states are visited with the same frequency}. 

The next step is to show that in a cycle, each node is 
exactly visited twice, i.e., all cycles of length
$4\cdot 2^N$, $6\cdot 2^N$, and so on, are in fact two,
three or more repetitions of a $2\cdot 2^N$-cycle. Again,
assume the system is moving on a cycle. If this cycle were truly
longer than $2\cdot 2^N$, there must, at the point $t=2\cdot 2^N$, 
be nodes that have been visited three or more times 
while others have not been visited for the second time --
the visit numbers must add up to $2\cdot2^N$, and if all
visit numbers were equal to 2, the cycle would be complete.
More specifically, there must be an earlier time when 
all visit numbers $v^{\nu}_t$ are either 0, 1, or 2, 
and one node is about to be visited for the third time.
It suffices to show that this cannot happen to prove that 
the cycle cannot be longer than $2\cdot 2^N$.

The third visit to a node $\mu$ with $v^{\mu}=2$ cannot come
from a predecessor (let us say, $^0\mu$) with a visit number
of $v^{^0\mu}=0$, for the obvious reason that this
predecessor has not been visited yet. It also cannot come
from a predecessor with $v^{^0\mu} =1$: if $v^{\mu}=2$, 
either it must have been visited before from $^0\mu$
(which it cannot -- the predecessor has only had one
visit so far), or it must have been reached twice from
$^1\mu$ -- this is impossible as well, since it means that
$v^{^1\mu}\geq 3$. For similar reasons, we can exclude a
visit from a node with $v^{^0\mu}=2$: either $v^{^1\mu}\geq
3$ as before, or the first visit to $^0\mu$ led to $\mu$ -- then the 
second cannot. This means that all nodes must receive two
visits -- thus finishing a cycle -- before one of them 
can be visited for the third time. The question arises why
this line of reasoning does not hold true during the
transient. The key lies in the observation that the first
node to receive three visits is the node where the system 
was started -- it did not get its first visit {\em from}
anywhere on the graph, so the arguments are not applicable.

Using all previous conclusions, the cycles turn out to be
solutions to a well-studied combinatorial problem, and the
number of different cycles can be found in the literature:
since, during a cycle, each string $\mu$ appears exactly once followed by 
each of ist successors, the cycle is a sequence in which
each $N+1$-bit pattern, i.e. each node on the DeBruijn 
graph of order $N+1$, appears exactly once. Such a 
sequence is a {\em Hamiltonian circuit}, also known as a
{\em full cycle}, on the $N+1$-graph. For a review on the
properties of these cycles, consult
Ref. \cite{Fredricksen:Survey}. One of the earliest and most
central results on the topic is the number
of different cycles, which is $2^{2^N - (N+1)}$\cite{Flye:48}. 

All possible full cycles on the $N+1$-graph can be generated by the
antipersistent walk on the $N$-graph: write down the
desired sequence starting at some arbitrary point, look for the first
occurrence of each $N$-bit pattern $\mu$, and set the
corresponding table entry to the bit that follows it. 
Starting the antipersistent walk at the first pattern of 
the  desired sequence, the antipersistent walk will
reproduce it. 

Since all cycles are of length $2\cdot 2^N$,
a total of $2\cdot 2^N \times 2^{2^N - (N+1)}= 2^{(2^N)}$ 
states is part of a cycle.
As mentioned before, the total number of possible states is 
$\Omega = 2^N \cdot 2^{(2^N)}$, which means that a fraction 
of $2^{(2^N)}/\Omega = 2^{-N}$ of possible states is part 
of a cycle. 

\section{Properties of the transient}
\label{AP-proptrans}
The length $\tau$ of the transient is the number of steps 
taken before the system enters a cycle.
The distribution of transient lengths is less accessible to
analytical approaches, but easy to measure in computer 
simulations, either by complete enumeration for small
systems or by Monte Carlo for larger ones. 
The following picture emerges, as seen in
Fig. \ref{AP-trans}: 

\begin{figure}
  \epsfxsize= 1.0\columnwidth
  \epsffile{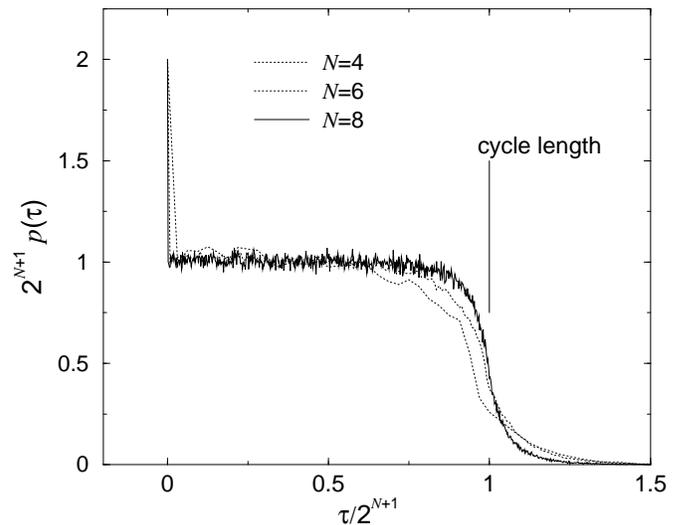}
  \caption{Distribution of transient lengths $\tau$, 
rescaled by the cycle length $2^{N+1}$.} 
  \label{AP-trans}
\end{figure}

The probabilty
for transient length $\tau=0$ is just the probability of 
hitting a cycle right away, and thus the fraction of state
space filled with cycles. As mentioned, this is equal to 
$2^{-N}$. 

The probability distribution is more or less flat 
for $1\leq \tau \leq 2^{N+1}$, the cycle length. From
normalization constraints, it follows that $p(\tau)\approx
2^{-(N+1)}$ in that range.

Near $\tau=2^{N+1}$, there is an exponential drop
reminiscent of a phase transition, which gets steeper with
increasing $N$. Even for small $N$, no transients longer
than $2^{N+2}$ have been observed.

\section{Antipersistence on different timescales}
\label{AP-time}
Consider the following prediction algorithm: an
observer looks at $N_{obs}$-bit strings from a binary 
time series, writes down the bit that followed the 
pattern in the appropriate entry of his decision table 
(which, of course, has $2^{N_{obs}}$ rows),
and predicts that when that pattern occurs the next time,
it will be followed by that same bit written in his table.

This algorithm, which could be labelled ``blind reliance on 
recent experience'',
works well for persistent time series, and it is similar in
spirit to what we all do instinctively -- similar situations
usually lead to similar consequences. An interesting
quantity is the success rate $s(N, N_{obs})$ of this prediction
algorithm when it predicts a completely antipersistent walk
on the $N$-graph. For simplicity's sake, we consider the
long-time limit, in which both the generator and the
observer move on a cycle.

If the observer looks at the same time window as the
generator ($N_{obs}=N$), it is obvious that the success
rate will be $0$ -- since each pattern is continued with 
alternating bits on each visit. In that sense, the generated
time series is {\em anti-predictable} for this specific
prediction algorithm (for more on anti-predictable
sequences, see \cite{Kinzel:Seq.,Metzler:CBG}).

For an observer with
a slightly  larger window, the picture changes: as mentioned
above, the antipersistent cycle corresponds to a
Hamiltionian cycle on the $N+1$-graph, which is completely
persistent and predictable with $100 \%$ accuracy. For even
larger $N_{obs}$, the antipersistent cycle looks like a
closed path which includes only a fraction of
$2^{N-(N_{obs}+1)}$ of nodes on the $N_{obs}$-graph.
Prediction is again $100\%$ reliable, and the observer does
not even need all of his storage capacity to handle the
cases that occur.

\begin{figure}
\centering 
  \epsfxsize= 0.7\columnwidth
  \epsffile{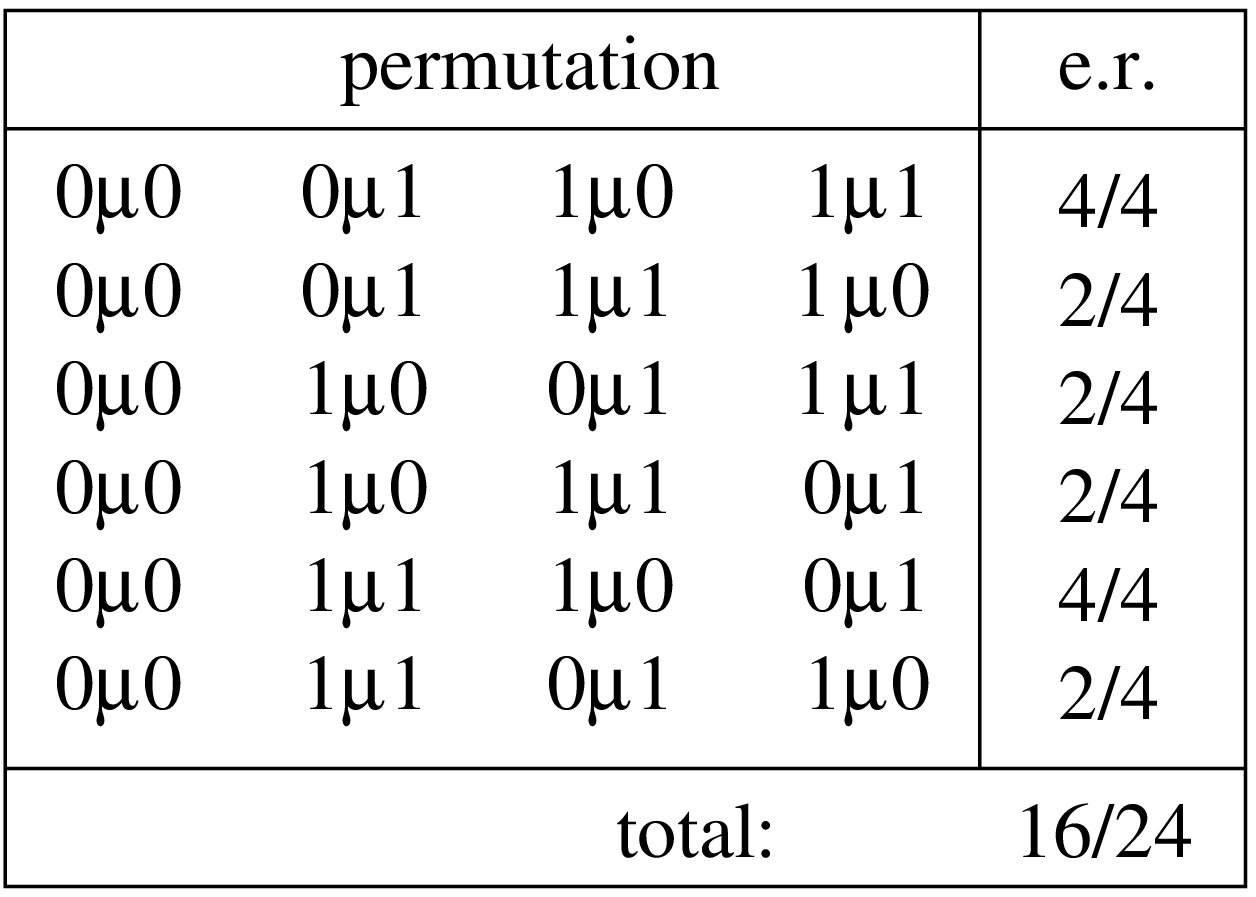}

        \caption{Possible permutations of predecessors and 
successors of a pattern $\mu$ on a cycle. The error rate,
given in the last column, is the rate of flips between 
0 and 1 in the sequence of successors.}
        \label{AP-tab1}
\end{figure}

If the observer has a shorter time window than the
generator, more than one of the generator's patterns will 
affect the same table entry for the observer. For example,
an $N-1$-bit pattern $\nu$ corresponds to either of the
$N$-bit strings $0\nu$ or $1\nu$, both of which occur twice
in the $N$-cycle, each time followed by a different
successor. The success rate of the predictor depends on the
sequence in which these combinations occur; if each
permutation of $0\nu 0$, $0\nu 1$, $1\nu 0$ and $1\nu 1$
has the same probability, the success rate for all patterns
is the average over the different permutations. Figure
\ref{AP-tab1} shows that this average is $1/3$ for 
$N_{obs}=N-1$.

For $N_{obs}=N-2$, all permutations of 
eight combinations of predecessors and successors have to be
taken into account -- a task best left to computer algebra 
programs, which yield $\langle s(N,N-1)\rangle =  3/7$, in excellent 
agreement with simulations (see Fig. \ref{AP-predict}. 
Larger differences in the time
window are beyond even the scope of computer programs; 
however, it can be argued that for larger $N-N_{obs}$,
the visits to the $N_{obs}$-nodes become more and more
random, and $s(N,N_{obs})$ will tend to $1/2$.

\section{Stochastic antipersistence}
\label{AP-stoch}
All of these observations relied on the fact that the
generator is on a cycle with well-known properties. It is 
thus interesting to ask how stable these results are if the
sequence is not completely antipersistent. The simplest 
generalization is to introduce a probabilty $p$ for 
changing the table entry/ exit when visiting a node:
$p=1$ reproduces the completely antipersistent walk; 
$p=0$ is equivalent to using a constant (quenched)
decision table, and $p=1/2$ generates a completely random
time series. 

\begin{figure}[t]
  \epsfxsize= 1.0\columnwidth
  \epsffile{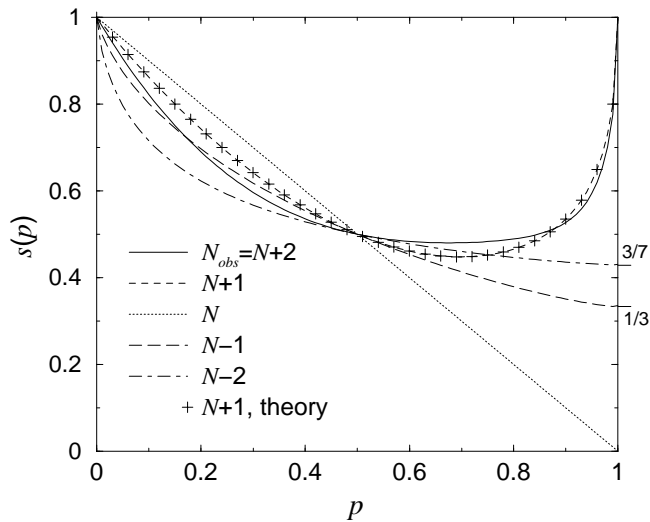}
  \caption{Success rate of an observer keeping a table of
    the recent occurrence of $N_{obs}$-bit strings and the
    respective following bit, for $N=16$. $p$ is the
    probability of flipping the exit in the generating
    graph. The symbols labelled ``theory'' were calculated 
    using Eq. (\ref{AP-s_p}) with approximate probabilities
    $\pi_l(p,N)$ taken from the simulation itself.} 
  \label{AP-predict}
\end{figure}

A first intuitive guess would be that even a small deviation
from deterministic dynamics completely destroys all predictability:
after all, on a path of length $2^{N+1}$, there are
on the average $(1-p)2^{N+1}$ occasions where the the
sequence is continued persistently, thus leaving the cycle.
Indeed, a single ``error'' is usually enough to move the 
system from one cycle to another; however, much of the
local structure remains untouched.
It turns out that the functions $s(N, N_{obs},p)$ of
prediction rates converge for large $N$ (meaning roughly
$N>12$) to a set of curves that depend only on $p$ and $N-
N_{obs}$, which is displayed in Fig. \ref{AP-predict}.

The limit values for $p=1$ have been explained
above, and they are approached continuosly for
$p\rightarrow 1$. For $p=0.5$, the curves intersect at $s=0.5$ -- no
prediction beyond guessing is possible. For small $p$, 
all curves converge to 1: the system is dominated by 
short loops in which only a small fraction of the possible
states participate, and those are predicted with high accuracy.

Interestingly, between $p=0.5$ and roughly $p=0.85$, 
all shown curves are below 0.5, meaning that even observers
with longer memory predict the sequence with less than 
$50\%$ accuracy.
I will give an analytical argument why this is the case
for $N_{obs}=N+1$. An $N+1$-bit pattern $\nu$ is a
combination of an $N$-bit pattern $\mu$ and one of its 
predecessors, let us say $^0\mu$, whereas the companion
state $\hat{\nu}$ is a combination of $\mu$ and the 
other predecessor $^1\mu$. A visit to either $\nu$ or
$\hat{\nu}$ switches the exit of $\mu$ with probability $p$.
Consider two subsequent visits to $\nu$, with some number 
$l$ of visits to $\hat{\nu}$ between them. The probability
$s(p, N, N+1)$ of continuing with the same bit after these
two visits is a sum of two probabilities:
either the exit of $\mu$ was switched upon leaving $\nu$
the first time and then switched an odd number of times
during the $l$ visits to $\hat{\nu}$, or it was not
switched the first time and switched an even number of
times in between. Given $p$ and the probability 
$\pi_l(p,N)$ of having $l$ intermediate visits to $\hat{\nu}$, one then
obtains by basic combinatorics
\be
s(p,N, N+1) = \sum_{l=0}^{\infty} \frac{1}{2} \pi_l(p,N) 
[1+ (1-2p)^{l+1}].
\label{AP-s_p}
\ee  
Unfortunately, $\pi_l(p,N)$ does not seem to be 
analytically accessible for general $p$. It can be measured
in simulations, and the accuracy of Eq. (\ref{AP-s_p})
verified (see Fig. \ref{AP-predict}); also, for $p=1/2$, 
since the system does a completely random walk on the
graph, one gets the simple distribution $\pi_l(1/2,N) =
2^{-(l+1)}$. Assuming that this distribution does not 
change discontinuously near $p=1/2$, Eq. (\ref{AP-s_p})
yields the  approximation $s(1/2+\delta p, N, N+1) \approx
1/(2+2\delta p)
\approx (1/2)(1-\delta p)$. This is obviously $<1/2$ for
$\delta p>0$, i.e., $p>1/2$.

 \ \\
The error rate $1-s(p,N,N_{obs})$ is a measure of the
antipersistence of the time series on the scale $N_{obs}$ --
for $N_{obs}=N$, $1-s$ is completely equivalent to the
antipersistence parameter $p$ of the underlying dynamics.
However, even if $s(p, N, N+1)$ could be calculated, 
it would not be possible to recursively apply this function
to find the antipersistence on the scales $N+2, N+3$ etc.
In other words, 
\bea
1-s(p, N, N+1) &\neq& \nonumber \\
 1-s( \{1-s(p, N, N+1)\},N+1, N+2).
\eea
It is thus not sufficient to give a single parameter
$p$, or $1-s$, for some $N$ in order to characterize 
the behaviour of a time series completely and to 
calculate its predictabilty on other scales of 
observation. The scale on which the dynamics work is 
important as well.  

\section{Results and conclusion}
I introduced a deterministic algorithm that generates a 
binary time series that is completely antipersistent with
respect to strings of length $N$. After a short
transient, the algorithm runs into cycles of length
$2\cdot2^N$, in which each string appears exactly twice. 
These cycles correspond to Hamiltonian paths on a 
DeBruijn graph of order $N+1$. 

The cycle length is much larger than the typical cycle
length of a graph with fixed decision tables. This seems
typical for antipredictable sequences: sequences that
can be predicted with $100\%$ accuracy by some prediction
algorithm usually do not require adaptation of the algorithm's
parameters, whereas antipredictable sequences explore the
combined phase space of the sequence and the generating 
algorithm, allowing for more, longer, and more complex cycles. 
In this case, however, the dynamics allow for fairly simple
proofs of the properties of the cycles.

Observers that keep track of the most recent occurrence
of $N_{obs}$-bit strings can predict the completely 
antipersistent cycle with $100\%$ accuracy if $N_{obs}>N$,
and with less than $50\%$ success rate if $N_{obs}\leq N$.
If the stochasticity is introduced by means of a probability
$p$ of  flipping the exit edges, the success rate even of
observers with $N_{obs}>N$ can drop below $50\%$,
which shows that larger memory does
not necessarily give better results.  The rate
of antipersistence on one scale is not sufficient to
calculate the rate for other scales.

\section{Acknowledgment}
I am grateful for discussions with and helpful ideas from
Hanan Rosemarin, Andreas Engel, Stephan Mertens, Ido Kanter,
Michael Biehl, and Wolfgang Kinzel, and for financial support by the
German-Israeli Foundation.

\end{multicols}

\end{document}